\documentclass[aip,apl,reprint]{revtex4-1}%
\usepackage{amsmath}
\usepackage{amssymb}
\usepackage{graphicx}
\usepackage{amsfonts}
\usepackage{bm}
\usepackage{color}

\begin{document}
\title{Electric-field control of collective spin excitations in N\'{e}el-type skyrmions}
\author{Hong-Bo Chen }
\affiliation{Ningbo Institute of Technology, Zhejiang University, Ningbo 315100, China }
\author{You-Quan Li}
\affiliation{Zhejiang Institute of Modern Physics and Department of Physics, Zhejiang
University, Hangzhou 310027, China }
\affiliation{Collaborative Innovation Center of Advanced Microstructures, Nanjing
University, Nanjing 210093, China}

\begin{abstract}
We demonstrate that an electric field could activate the three
low-energy eigenmodes of the N\'{e}el-type skyrmion lattice via the electrically induced Dzyaloshinskii-Moriya interaction.
In particular, we predict that the relative intensity of the clockwise rotation mode against the counter-clockwise rotation mode is significantly enhanced for the electrical activation in comparison with the magnetic activation. We also discover that
the electrically and magnetically active modes obey unique selection rules.
These findings promise a fresh pathway towards
energy-efficient electrical manipulation of skyrmion excitations for future
skyrmion-based magnonics.

\end{abstract}
\maketitle

%\date{\today }

Magnetic skyrmions are nanoscale, topologically stable spin
textures.\cite{Pfleiderer2006} Because of their unique feature of topology and
controllable manipulation, they are in the focus of current research and an
emerging area for building highly efficient next-generation spintronic
devices.\cite{Nagaosa2013,Li2015,Fert2017review,Klaui2018} Three distinct
classes of magnetic skyrmions have by now been experimentally discovered,
namely, Bloch skyrmion,\cite{Science2009,Tokura2010,Seki2012} N\'{e}el
skyrmion,\cite{Neel2015,Heinze2011} and antiskyrmion,\cite{AntiSky2017} which
can be stabilized by the bulk, interfacial, and anisotropic
Dzyaloshinskii-Moriya interaction (DMI),\cite{DMI} respectively. Controlling
of magnetic skyrmions is an essential issue regarding their potential
application in devices. Nowadays, various schemes have been demonstrated such
as, the use of electric
currents,\cite{current1,current2,current3,Li2011prb,Liu2013jpc,current4}
temperature gradient,\cite{temp1,temp2,temp3}
microwaves,\cite{microwave2015,microwave2018} or
strain.\cite{strain} In particular, manipulating magnetic skyrmions by an
electric field has been achieved through the voltage-controlled magnetic
anisotropy in ultrathin metal
films\cite{EF-PMA3,EF-PMA5}, or the magnetoelectric
coupling in multiferroic insulator.\cite{EF2,EF-ME2015} Moreover, {a very
intriguing electric-field induced DMI mechanism has recently been
demonstrated\cite{EF-DMI,EF-DMI2} to control the skyrmion dynamics}.

Meanwhile, intense activity has been devoted to the collective spin
excitations of skyrmions, which is promising for the potential application in
the field of magnonics.\cite{SW-review1,SW-review2} The fundamental eigenmodes
of skyrmions can be excited usually by an ac magnetic field. In the skyrmion
crystal phase, three distinctive low-energy eigenmodes, i.e., the breathing
(BR) mode, the clockwise (CW), and the counter-clockwise (CCW) rotation modes,
have been first revealed theoretically by Mochizuki\cite{BlochSW2012} and
subsequently demonstrated experimentally in both
Bloch\cite{BlochSW2012-2,BlochSW2013,BlochSW2015} and N\'{e}%
el\cite{NeelSW2016,NeelSW2019} skyrmions. The two rotational modes CW and CCW
are characterized by the core of skyrmion rotates simultaneously clockwise and
counter-clockwise, respectively, under an in-plane ac magnetic field, while
the breathing mode is excited by an out-of-plane ac field with the skyrmion
expanding and shrinking coherently. Currently, skyrmion excitations have been
examined primarily by means of microwave magnetic fields. Therefore, seeking
mechanisms that enable electric-field controlled skyrmion excitations are
particularly attractive since electric field is much easier to manipulate than
magnetic field. A natural way is to use the magnetoelectric coupling in
insulator, which has recently achieved in a multiferroic Bloch-type skyrmion
Cu$_{2}$OSeO$_{3}$.\cite{Mochizuki-2013,Liu2013} However, for widely existing
N\'{e}el-type skyrmions, this has rarely been explored before and remains an
outstanding theoretical issue.

In this work we aim at exploring the effect that an electric field can have on
the collective spin excitations in the N\'{e}el-type skyrmions. To study this,
we employ a novel electric-field induced DMI mechanism developed
recently.\cite{DMI-E,DMI-E-1,DMI-E-3} This coupling can also be
considered as the spin flexoelectric effect.\cite{DMI-E-5} We demonstrate the
electrical activation of spin excitations in the N\'{e}el skyrmion lattice and
determine the spectrum of three eigenmodes. We predict that the
higher-lying CW rotation mode, absent in the previous conventional microwave magnetic field
experiments, would be easily observed in the microwave electric field experiment
because its relatively strong spectral weight driven by the ac electric field.
Additionally, we discover that the same skyrmion-mode excited simultaneously
by the electric and magnetic field obeys the identical selection rule, which is
totally different from that of the Bloch-type skyrmions.

We consider a classical spin model on a two-dimensional square lattice in the
$x$-$y$ plane. The effective Hamiltonian supporting the N\'{e}el skyrmion is
given by
\begin{equation}
H_{0}=-\sum_{\left\langle i,j\right\rangle }\big[  J\mathbf{S}_{i}%
\cdot\mathbf{S}_{j}+\mathbf{D}_{ij}\cdot(\mathbf{S}_{i}\times\mathbf{S}%
_{j})\big]  -\sum_{i}\mathbf{B}\cdot\mathbf{S}_{i},\label{eq:H1}%
\end{equation}
where $\mathbf{S}_{i}$ is the normalised spin at the site $i$, and the
summation $\langle i,j\rangle$ runs over all pairs of nearest-neighbour spins.
$J>0$ is the ferromagnetic exchange coupling, $\mathbf{B}$ is the external
magnetic field for $\mathbf{B}=B\hat{z}$, and $\mathbf{D}_{ij}=$ $D\hat
{z}\times\mathbf{\hat{r}}_{ij}$ is the intrinsic interfacial DMI with
$\mathbf{\hat{r}}_{ij}$ the unit vector from spins $i$ to $j$, the DM constant
$D=J\tan(2\pi/p)$ defining the pitch length ($p$) of the modulated spin
structures. In what follows, we fix the exchange constant $J=1$ for the energy
units, and the DM value with $D/J=0.445$ corresponding to the pitch length
$p=15$. The calculations were performed on a $66\times66$ lattices with
periodic boundary conditions imposed along $x$- and $y$- direction.\\

\begin{figure}[t]
\includegraphics[width=8cm]{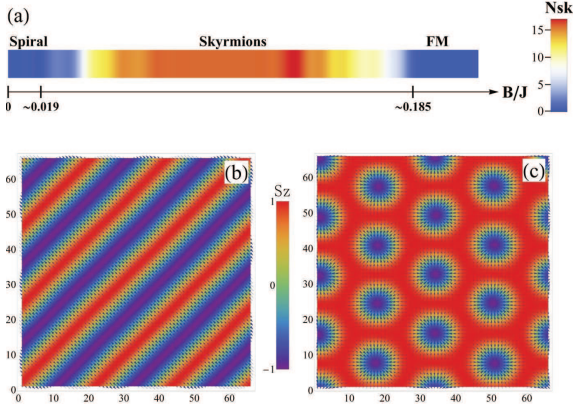}
\caption{(color online) (a) Ground-state phase diagram of the model (\ref{eq:H1}) as a function of the external magnetic field $\mathbf{B}=B\hat{z}$. Three distinct spin phases, spiral state, skyrmions state and ferromagnetic (FM) state are appeared. The color depicts the calculated cumulative skyrmion number $N_{\rm{sk}}$.
(b),(c) Representative ground-state spin textures of the system: (b) spiral state for $B/J=0.0$, (c) skyrmion lattice state for $B/J=0.11$. The arrows represent the in-plane components of spins and the color indicates their $z$ components. }
\label{Fig:fig1}
\end{figure}

We computed the time evolution of the spin dynamics by solving the
Landau-Lifshitz-Gilbert (LLG) equation, which is written as\cite{BlochSW2012}%
\begin{equation}
\frac{\partial\mathbf{S}_{i}}{\partial t}=-\frac{1}{1+\alpha^{2}}\left[
\mathbf{S}_{i}\times\mathbf{H}_{i}^{\mathrm{eff}}+\frac{\alpha}{S}%
\mathbf{S}_{i}\times(\mathbf{S}_{i}\times\mathbf{H}_{i}^{\mathrm{eff}%
})\right]  , \label{eq:LLG}%
\end{equation}
where $\alpha$ is the Gilbert-damping parameter, and $\mathbf{H}%
_{i}^{\mathrm{eff}}=-\partial H/\partial\mathbf{S}_{i}$ is the effective local
field acting on the $i$th spin $\mathbf{S}_{i}$. $H=H_{0}+H_{\mathrm{E}}(t)$, $H_{0}$ is the spin model Hamiltonian given by Eq. (\ref{eq:H1}), while
the term $H_{\mathrm{E}}(t)=\gamma(\mathbf{E}(t)\times\mathbf{\hat{r}}_{ij}%
)\cdot(\mathbf{S}_{i}\times\mathbf{S}_{j})$ represents the electric-field
induced DMI,\cite{DMI-E,DMI-E-1,DMI-E-3,Muchizuki2018} with the coupling
constant $\gamma =Jea/E_{\mathrm{SO}}$, where $e$ is the electron
charge, $a$ is the lattice constant, and $E_{\mathrm{%
SO}}$ is an energy scale associated with the inverse of the strength of the spin-orbit coupling. Given the typical parameter values, $J\sim1 $meV, $a\sim 10${\AA}, and $E_{\mathrm{%
SO}}\sim 1$eV, we roughly estimate $\gamma \sim 10^{-31}$ Cm. Each classical spin $\mathbf{S}_{i}$ is taken to have the unit
length, $S=1$. For the value of $\alpha$, we adopt $\alpha=0.04$ and $0.005$
for simulations of the phase diagram and the dynamic response of system to the applied
ac fields, respectively. We use the fourth-order Runge-Kutta method for numerical
integration of the LLG equation (\ref{eq:LLG}).

Figure \ref{Fig:fig1} shows the phase diagram at zero temperature of
the spin model (\ref{eq:H1}) as a function of magnetic field $B$, and two
representative spin textures of the phases. The phase diagram is obtained from
the combination of the classical Monte-Carlo simulations and the LLG equation.
We first start with the Monte-Carlo simulated annealing for the model
(\ref{eq:H1}) to obtain stable spin states until a low temperature
($k_{B}T/J=0.01$) is reached. Then, using the LLG equation, we further
relax them by sufficient time evolution to obtain its ground-state. The
skyrmions state in the system can be recognised effectively by a nonzero
skyrmion number (topological charge), which is given by%
\begin{equation}
N_{\mathrm{sk}}=\frac{1}{4\pi}\int dxdy\ \mathbf{S}\cdot\left(  \partial
_{x}\mathbf{S}\times\partial_{y}\mathbf{S}\right)  . \label{Eq:Q}%
\end{equation}
In practice for a square lattice, we compute the discretized version of this
expression.\cite{Nsk-1,Nsk-2} Figure \ref{Fig:fig1}(a) reveals that three
different types of spin states, specifically, spiral state, skyrmions state,
and the ferromagnetic (FM) state successively emerge as $B$ increasing. The
colors in the phase diagram indicate the calculated total skyrmion number,
which is a measure for the total number of skyrmions in the system. We note
that the spiral and FM phases have a zero skyrmion number. The phase boundary
between the spiral phase and the skyrmions phase is at $B/J\sim0.019$, and
that between the skyrmions and the ferromagnetic phase is at $B/J\sim0.185$.
Figures \ref{Fig:fig1}(b) and (c) display the representative spin textures of
spiral and skyrmion lattice phases for $B/J=0.0$ and $0.11$, respectively.
The skyrmion lattice phase shown in Fig. \ref{Fig:fig1}\ (c) will be used to
simulate the spin excitations in the follows.

\begin{figure}
[b]\includegraphics[width=8.3cm]{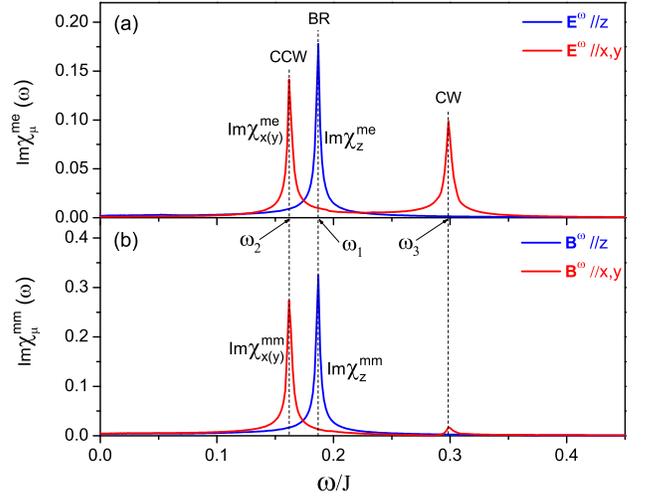}
\caption{(color online) (a) Imaginary parts of the dynamical electromagnetic susceptibilities $\rm{Im}\chi_{\mu}^{\rm{me}}(\omega)$ ($\mu=x,y,z$) obtained after applying a $\delta$-function electric field pulse $\mathbf{E}(t)=\delta(t)\mathbf{E}^{\omega}$ to the stable skyrmion lattice state [Fig. \ref{Fig:fig1}(c)], where $\rm{Im}\chi_{x(y)}^{\rm{me}}(\omega)$ and $\rm{Im}\chi_{z}^{\rm{me}}(\omega)$ for $\mathbf{E}^{\omega}\Vert \hat{x}(\hat{y})$ and $\mathbf{E}^{\omega}\Vert \hat{z}$, respectively. The three resonant modes were ascribed to the BR, CCW, and CW mode, which are located at $\omega_{1}/J=0.1869$, $\omega_{2}/J=0.1618$, and $\omega_{3}/J=0.2985$, respectively. (b) The imaginary parts of the dynamical magnetic susceptibilities, $\rm{Im}\chi_{x(y)}^{\rm{mm}}(\omega)$ and $\rm{Im}\chi_{z}^{\rm{mm}}(\omega)$ for $\mathbf{B}^{\omega}\Vert \hat{x}(\hat{y})$ and $\mathbf{B}^{\omega}\Vert \hat{z}$, respectively. Comparison between the spectra of $\rm{Im}\chi_{\mu}^{\rm{me}}(\omega)$ and $\rm{Im}\chi_{\mu}^{\rm{mm}}(\omega)$, the electrically and magnetically active resonances are located at the identical frequencies with the same selection rules. }
\label{Fig:fig2}
\end{figure}

To study the spin excitations of the system, we calculate the dynamical susceptibilities, which
are defined as\cite{BlochSW2012}%
\begin{equation}
\chi_{\mu}^{\mathrm{{mm}}}(\omega)=\frac{\Delta S_{\mu}(\omega)}{B_{\mu
}(\omega)},\ \chi_{\mu}^{\mathrm{{me}}}(\omega)=\frac{\Delta S_{\mu}(\omega
)}{E_{\mu}(\omega)},
\end{equation}
where the subscript $\mu$ stands for $x$, $y$ or $z$. $B_{\mu}(\omega)$ and
$E_{\mu}(\omega)$ are the Fourier transform of the time-dependent pulse of
magnetic field $B_{\mu}(t)$ and electric field $E_{\mu}(t)$ applied in the
$\mu$ direction, respectively. Here, we employ a time-localized, uniform
$\delta$-function pulse, i.e., $\mathbf{E}(t)=\delta(t)\mathbf{E}^{\omega}$
applied at $t=0$. And $\Delta S_{\mu}(\omega)$ is the Fourier transform of the
$\mu$-component of the spatially averaged spin $\Delta\mathbf{S}%
(t)=\mathbf{S}(t)-\mathbf{S}(0)$, with $\mathbf{S}(t)=(1/N)\sum_{i=1}%
\mathbf{S}_{i}(t)$, which is the transient response of the system under the
intense pulse of external fields. In the LLG simulation, we use a smaller
value of the damping constant, $\alpha=0.005$, which is allowed for better
frequency resolution of the excited modes.

Figure 2 shows the imaginary parts of the dynamical susceptibilities for the
selected skyrmion lattice (Fig. \ref{Fig:fig1} c) under an in-plane and
out-of-plane $\delta$-function pulse of electric and magnetic field. In Fig.
\ref{Fig:fig2}(a), we firstly display the imaginary parts of the dynamical
electromagnetic susceptibilities $\operatorname{Im}\chi_{\mu}^{\mathrm{me}%
}(\omega)$, with $\operatorname{Im}\chi_{z}^{\mathrm{me}}(\omega)$ and
$\operatorname{Im}\chi_{x(y)}^{\mathrm{me}}(\omega)$ for the out-of-plane ac
electric field $\mathbf{E}^{\omega}\Vert\hat{z}$ and in-plane ac electric
field $\mathbf{E}^{\omega}\Vert\hat{x}(\hat{y})$, respectively. We can see that
three resonant peaks are clearly exhibited. In $\operatorname{Im}\chi
_{z}^{\mathrm{me}}(\omega)$ for $\mathbf{E}^{\omega}\Vert\hat{z}$, we observe
a single resonant peak centered at $\omega_{1}/J=0.1869$. As will be justified
in detail below in Fig. \ref{Fig:fig3} (a), this resonant mode can be
identified as the so-called breathing mode, where all the skyrmions in the
skyrmion lattice periodically shrink and expand in a uniform way under the
out-of-plane driving electric field. We should note that this electrically
activated breathing mode has also been identified very
recently.\cite{Muchizuki2018} Furthermore, in $\operatorname{Im}\chi
_{x(y)}^{\mathrm{me}}$ for $\mathbf{E}^{\omega}\Vert\hat{x}(\hat{y})$, we see
two distinct resonant peaks, a lower-lying peak located at $\omega
_{2}/J=0.1618$, and a higher-lying peak at $\omega_{3}/J=0.2985$. In fact, as
will be shown in Figs. \ref{Fig:fig3}(b) and \ref{Fig:fig3}(c), the
lower-lying mode ($\omega_{2}$) can be assigned to the CCW rotation mode,
while the higher-lying mode ($\omega_{3}$) to the CW rotation mode, where all
skyrmion cores rotate simultaneously counter-clockwise or clockwise under the in-plane
driving electric field.

For comparison, we also present in Fig. \ref{Fig:fig2}(b) the imaginary parts
of the calculated magnetic susceptibilities, $\operatorname{Im}\chi
_{x(y)}^{\mathrm{mm}}(\omega)$ and $\operatorname{Im}\chi_{z}^{\mathrm{mm}%
}(\omega)$ for the in-plane magnetic field $\mathbf{B}^{\omega}\Vert\hat
{x}(\hat{y})$ and the out-of-plane $\mathbf{B}^{\omega}\Vert\hat{z}$,
respectively. Three magnetically active resonance modes are also clearly
observed, which are consistent with previous experimental observations
\cite{NeelSW2016,NeelSW2019} as well as theoretical investigations
\cite{BlochSW2012,NeelSW2017}. Compared to the spectra of $\operatorname{Im}%
\chi_{\mu}^{\mathrm{me}}(\omega)$ in Fig. \ref{Fig:fig2}(a), one can find that
three electrically active resonances can be seen in the spectra of
corresponding magnetic susceptibilities $\operatorname{Im}\chi_{\mu
}^{\mathrm{mm}}(\omega)$ at the identical frequencies. This means that both
the electric and magnetic fields excite the same modes with the same selection
rules, which is in striking contrast to the case of multiferroic Cu$_{2}%
$OSeO$_{3}$ hosting Bloch-type skyrmions where the same skyrmon-mode active to
both the electric and magnetic field obeys a different selection
rule.\cite{Mochizuki-2013} Moreover, we should note that the
higher-lying CW mode was not easily detected experimentally via the microwaves
magnetic field\cite{BlochSW2012-2,BlochSW2013,NeelSW2019} due to its rather weak spectra
intensity according to the theoretical calculations.\cite{BlochSW2012,NeelSW2019}
However, one can find in Fig. \ref{Fig:fig2}(a) that the relative intensity of the CW
mode against the CCW mode in $\operatorname{Im}\chi_{x(y)}^{\mathrm{me}}$ is significantly enhanced
under the ac electric fields. Therefore, our simulations indicate that the CW
mode is expected to be identified experimentally through the
microwave electric field.

\begin{figure}[t]
\includegraphics[width=8.5cm]{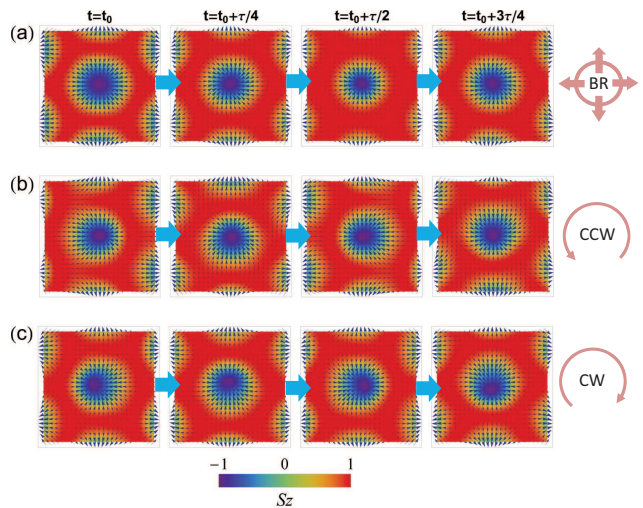}
\caption{(color online) Time evolution
snapshots of the spatial spin profiles for a constituent skyrmion of the
skyrmion lattice phase [Fig. \ref{Fig:fig1}(c)] induced by an oscillating
sinusoidal electric field, with a time increment of $1/4$ of the oscillation
period $\tau$. The arrows denote the in-plane spin
components, while the contour plot represents the out-of-plane $z$ component.
(a) The breathing mode excited by an out-of-plane oscillating electric field
$\mathbf{E}(t)=E^{\omega}_{z}\sin(\omega_{1}t)\hat{z}$ with field amplitude $\gamma E^{\omega}%
_{z}=0.006$. (b) The CCW rotation mode excited by an in-plane oscillating electric
field $\mathbf{E}(t)=E^{\omega}_{x}\sin(\omega_{2}t)\hat{x}$ with $\gamma E^{\omega
}_{x}=0.015$. (c) The CW rotation mode excited by $\mathbf{E}(t)=E^{\omega
}_{x}\sin(\omega_{3}t)\hat{x}$ with $\gamma E^{\omega}_{x}=0.015$. }
\label{Fig:fig3}
\end{figure}

To exclusively classify each electrically active modes revealed in the spectra
of Fig. \ref{Fig:fig2}(a), we examine the spin dynamics of the skyrmion
lattice under a spatially uniform sinusoidal electric field $\mathbf{E}%
(t)=\mathbf{E}^{\omega}\sin(\omega_{i}t)$, where $\omega_{i}$ ($i$ = $1$-$3$)
are the corresponding eigenfrequencies of the skyrmion excitations, by
numerically solving the LLG equation. The Gilbert-damping constant is fixed at
$\alpha=0.005$. Since for each mode, all the skyrmions in the skyrmion lattice
behaves uniformly the same way, we focus on a constituent skyrmion. The
selected constituent skyrmion is between the sites $22\leq i_{x}\leq47$, and
$20\leq i_{y}\leq45$ shown in Fig. \ref{Fig:fig1}(c).

In Fig. \ref{Fig:fig3}, we depict four time representative snapshots of the
simulated spin texture evolution of a constituent skyrmion for three
electrically active eigenmodes, over one oscillation period $\tau$. The characteristic
motions of the three eigenmodes are clearly illustrated in the respective time
evolutions of their dynamic spatial spin profiles. In Fig. \ref{Fig:fig3}(a), we show the simulated spin dynamics of a
constituent skyrmion excited by an out-of-plane oscillating electric field
$\mathbf{E}(t)=E_{z}^{\omega}\sin(\omega_{1}t)\hat{z}$ with field amplitude
$\gamma E_{z}^{\omega}=0.006$. The size of each constituent skyrmion in the lattice
performs an oscillatingly shrinking and expanding motion, so it can be
unambiguously assigned to the breathing mode of the skyrmion lattice. In Figs.
\ref{Fig:fig3}(b) and (c), we applied an in-plane oscillating electric field
$\mathbf{E}(t)=E_{x}^{\omega}\sin(\omega_{i}t)\hat{x}$ with field amplitude
$\gamma E_{x}^{\omega}=0.015$, and the resonant frequencies $\omega_{i}$ fixed for
the lower-lying mode $\omega_{2}$ and the higher-lying mode $\omega_{3}$,
respectively. In Fig. \ref{Fig:fig3}(b), we can see that the spin profiles rotate in the sense of counter-clockwise for the
lower-lying mode $\omega_{2}$, while in Fig. \ref{Fig:fig3}(c) for the
higher-lying mode $\omega_{3}$ the rotation sense is clockwise. Therefore, the
mode $\omega_{2}$ and mode $\omega_{3}$ can be assigned to the CCW and CW
rotation mode, respectively.

In conclusion, we have theoretically demonstrated the electric-field control
of collective spin excitations of the N\'{e}el-type skyrmion lattice phase,
driven by an electrically induced DMI effect. We successfully identified the
three eigenmodes of the N\'{e}el-type skyrmion lattice by using the electric field.
Furthermore, we revealed that the higher-lying CW rotation mode excited by the
electric field has a relatively strong spectral intensity. We also showed that the same
skyrmion-mode activated simultaneously by the electric and magnetic field
exhibits the same selection rule. These features are expected to bring in a new
insight to electrically activate spin excitations of magnetic skyrmions.

\begin{acknowledgments}
We acknowledge support by the National Key R \& D Program of China (No. 2017YFA0304304), NSFC (Nos. 11547102 and 11604294), Zhejiang
Provincial Natural Science Foundation of China (No. LY16A040008), and Ningbo
Natural Science Foundation (2015A610003).
\end{acknowledgments}

\end{document}